\documentclass[prb,a4paper,twocolumn,showpacs,superscriptaddress,preprintnumbers,amsmath,amssymb,floatfix]{revtex4} \newcommand{\myscale}{0.5}

\usepackage{graphicx}
\usepackage{dcolumn}
\usepackage{bm}
\usepackage{color}

\newcommand{\sr}{Sr$_{14-x}$Ca$_x$Cu$_{24}$O$_{41}$}
\newcommand{\sro}{Sr$_{14}$Cu$_{24}$O$_{41}$}

\newcommand{\parl}[2]{\mv{#1}\|\mv{#2}}

\newcommand{\RH}{$R_H$}

\newcommand{\mv}[1]{\mathbf{#1}}                
\newcommand{\vek}[1]{$\mv{#1}$}                 

\newcommand{\mva}{\mv{a}}   \newcommand{\va}{$\mva$}
\newcommand{\mvb}{\mv{b}}   \newcommand{\vb}{$\mvb$}
\newcommand{\mvc}{\mv{c}}   \newcommand{\vc}{$\mvc$}

\newcommand{\mja}{\parl{j}{a}}          \newcommand{\ja}{$\mja$}
         
\newcommand{\mjc}{\parl{j}{c}}    \newcommand{\jc}{$\mjc$}

\newcommand{\mBc}{\parl{B}{b}}    \newcommand{\Bb}{$\mBc$}

\newcommand{\rhoa}{\rho_a}

\newcommand{\rhoc}{\rho_c}

\newcommand{\plane}[2]{#1\text{--}#2}
      
\newcommand{\mabplane}{\plane{\mva}{\mvb}}      \newcommand{\abplane}{$\mabplane$}

\begin{document}

\title{Influence of doping on the Hall coefficient in
Sr$_{14-x}$Ca$_x$Cu$_{24}$O$_{41}$}

\author{E. Tafra}
    \email{etafra@phy.hr}
    \affiliation{Department of Physics, Faculty of Science, University of Zagreb, P.O. Box 331, HR-10002 Zagreb, Croatia}

\author{B. Korin-Hamzi\'c}
    \affiliation{Institute of Physics, P.O.Box 304, HR-10001 Zagreb, Croatia}

\author{M. Basleti\'c}
    \affiliation{Department of Physics, Faculty of Science, P.O.Box 331, HR-10002 Zagreb, Croatia}

\author{A. Hamzi\'c}
    \affiliation{Department of Physics, Faculty of Science, P.O.Box 331, HR-10002 Zagreb, Croatia}

\author{M. Dressel}
    \affiliation{1.~Physikalisches Institut, Universit\"{a}t Stuttgart, Pfaffenwaldring 57, D-70550 Stuttgart, Germany}

\author{J. Akimitsu}
    \affiliation{Department of Physics and Mathematics, Aoyama-Gakuin University, Sagamihara, Kanagawa 229-8558, Japan}

\date{\today}

\begin{abstract}
We present Hall-effect measurements of two-leg ladder compounds
Sr$_{14-x}$Ca$_x$Cu$_{24}$O$_{41}$ ($0\leq x \leq 11.5$) with the
aim to determine the number of carriers participating in dc
transport. Distribution of holes between the ladder and chain
subsystems is one of the crucial questions important for
understanding the physics of these compounds. Our Hall-effect and
resistivity measurements show typical semiconducting behavior for
$x < 11.5$. However, for $x=11.5$, the results are completely different, and the Hall coefficient and
resistivity behavior are qualitatively similar to that of high
temperature copper-oxide superconductors. We have determined the
effective number of carriers at room temperature and compared it
to the number of holes in the ladders obtained by other
experimental techniques. We propose that going from $x=0$ to
$x=11.5$ less than 1 hole per f.u.\ is added to the ladders
and is responsible for a pronounced change in resistivity with Ca
doping.
\end{abstract}


\maketitle

\section{Introduction}

\sr\ compounds are a part of the larger family
$A_{14}$Cu$_{24}$O$_{41}$ ($A$ = Sr, Ca, La, Y, \ldots) which are
considered as quasi-one-dimensional (Q-1D) copper oxides due to
their pronounced anisotropy. They have been investigated
intensively during the last years because of their fascinating
physical properties,\cite{Dagotto99,Vule06} their close relation
to cuprate superconductors\cite{DagottoHTC,hussey} and especially
after the superconductivity was achieved under high pressure in a
material heavily doped with Ca.\cite{Uehara96}

The  incommensurate crystal structures of these compounds consist
of planes of quasi-one dimensional CuO$_2$ chains stacked
alternately with planes of two-leg Cu$_2$O$_3$
ladders.\cite{siegrist,mccarron} The orientation of these chains
and ladders defines the crystallographical \vek{c} axis and are
alternately stacked along the \vek{b} axis separated by layers of
Sr(Ca). Chain and ladder spin subsystems in \sro\ interact weakly
along the \vek{a} axis. Consequently, the resistivity shows
remarkable anisotropy, indicative of a quasi one-dimensional
electronic state.\cite{motoyama} Materials with different Sr/Ca
composition are isostructural with only minor modifications of the
bond lengths and angles.\cite{isobe} All cuprate superconductors
found up to now contain square CuO$_2$ planes, whereas \sr\ is the
only known superconducting copper oxide without a square lattice.
An important feature of the superconductivity in \sr\ is that it
occurs by carrier doping in the low-dimensional antiferromagnetic
spin system. This feature is common to the CuO$_2$ plane.
Therefore, the evolution of the electronic structure upon hole
doping is one of the key issues for understanding
superconductivity.

\sr\ is intrinsically hole doped due to stoichiometric reasons. In
order for the formula unit to remain electrically neutral, the
average copper valency must be $+2.25$ instead of $+2$. Thus,
there is one hole per four Cu ions, or 6 per f.u.  There
are four formula units per approximate super-structure cell,
implying a hole density of $\sim 6 \times10^{21}\,$cm$^{-3}$. The
often accepted doped hole density refers to 1 hole for every 14
Cu's in the ladder and 5 holes for every 10 Cu's in the chain
(that are localized).

The Madelung potential calculations\cite{mizuno} have shown that
for the $x = 0$ compound holes are staying essentially in the
chains where their localization leads to an insulating behavior.
Upon Ca substitution, which does not change the total hole count,
holes are transferred from the chains to the ladders and the
longitudinal conductivity (along the \vek{c} axis) increases,
leading to Q-1D metallic properties. This
hole transfer can be caused by the reduction of the distance
between chains and (Sr,Ca) layers, which results in an enhancement
of the electrostatic potentials in the chains. The redistribution
of the holes among chains and ladders (depending on the Ca
content) is one of the most important factors which controls the
physical properties (in parti\-cular superconductivity) of these
spin-ladders. However, experimentally, the precise amount of hole
transfer is still under discussion, since different experimental
techniques have provided contradictory results.

From the optical data,  Osafune {\it et al.}\cite{osafune}
concluded that there is one hole per f.u.\ in the ladders
for $x = 0$, and 2.8 for $x = 11$. Polarization-dependent near-edge x-ray-absorption fine structure (NEXAFS), as the technique which can probe
the hole distribution between chains and ladders most directly,
shows a much smaller hole transfer from the chains into the
ladders induced by Ca-substitution.\cite{nucker} The results
predict that 0.8 holes per f.u.\ are found in the ladders
for $x=0$ and 1.1 for the $x = 12$ compound. Also, Gotoh et
al.\cite{gotoh} performed an x-ray diffraction study on \sro\
crystals. The bond-valence sum calculation based on their data has
indicated that only 0.5 holes per f.u.\ reside in the
ladders for the $x = 0$ compound. Nuclear magnetic resonance (NMR)
and nuclear quadrupole resonance (NQR) spectroscopy measurements
were also performed on these compounds with different
results.\cite{magishi,thurber} Recently Piskunov et
al.\cite{piskunov} reported results of a $^{63}$Cu and $^{17}$O
study on \sr\ for $x=0$ and $x=12$. They   have found that the
change in hole number is $n(x = 12) - n(x = 0) \approx 0.42$ holes
per formula unit in the  ladder, which is in a good agreement with
NEXAFS results. They  also reported that the hole distribution
changes with temperature and pressure, concluding that the
decrease in temperature reduces the hole number in the ladders by
transferring them partly back to the chains, while pressure seems
to induce an additional transfer of holes to the ladders. More
recently, the distribution of holes in \sr\ was revisited with
semi-empirical re-analysis of the x-ray absorption.\cite{rusydi} This interpretation of the x-ray absorption spectroscopy (XAS) data lead
to much larger ladder hole densities than previously suggested;
i.e., it was concluded that there are 2.8 hole per f.u.\ in
the ladders for $x = 0$, and 4.4 for $x = 11$.

The aim of this paper is to contribute to the undoubtedly still
open question about the amount of holes that participate in the dc
transport in \sr\ for $0\leq x \leq 11.5$, by studying the
temperature dependence of the Hall coefficient -- the long missing
basic experiment. In our work, we  emphasize the temperature
region around room temperature where the obtained estimate for the
effective number of carriers could be compared with the results
obtained by different experimental methods.

\section{Experiment}

The \sr\ ($0 \leq x \leq 11.5$) samples used in the present work
were single crystals, made from calcined sintered polycrystalline
rods by the floating zone recrystallization method. Samples were
$3-5$ mm long and had a cross-section of $0.2 - 0.4$\,mm$^{2}$. The
resistivity and Hall-effect data, that will be presented here, are
for \va\ and \vc\ axes. The \vc\ direction is the highest
conductivity direction (along ladders and chains), the \va\
direction with intermediate conductivity is perpendicular to \vc\
in the \va-\vc\ ladder plane, while the \vb\ direction (with the
lowest conductivity) is perpendicular to the \va-\vc\ plane. For
the \vc\ and \va\ axes resistivity measurements, the samples have
been cut from a long single crystal along the \vc\ and \va\ axes.

Electrical contacts to the sample  (three pairs of Hall contacts
and one pair of current contacts  on the sides of the crystal)
were obtained by using a special DuPont 6838 silver paste.  The
paste was first applied directly to the surface and heated for 1 h at 450$^\circ$C  in oxygen flow atmosphere, and then the
 30\,$\mu$m gold wires were pasted to the baked contact areas of
the sample.

The measurements were conducted in the temperature range
$4.2\,\mathrm{K}<T<300\,\mathrm{K}$. Due to the semiconducting
nature of the samples, the electrical resistances varied from
$\sim 1$~$\Omega$ to $\sim 100$~M$\Omega$; depending on the
resistance, either ac or dc techniques were used for the
measurements. For resistivity measurements two pairs of voltage contacts on each side of the sample were used.

The Hall effect was measured in  5 and 9 T magnetic fields. The
magnetic field was always along the \vb\ direction (\Bb\
geometry). For all the samples ($0 \leq x \leq 11.5$)   the
current was   \ja,  whereas  for $x=0$ and $x=11.5$, Hall data
were also taken for current along the \vc\ direction (this enabled
us to check the possible anisotropy in the Hall effect). The
latter geometry would have been a preferred choice, but for some
samples, on our disposal, it was not eligible (for accurate Hall
effect measurements) to measure in that geometry due to unnested
voltages greater than 20\% (which are an indication of
inhomogeneous current flow). To decrease the inhomogeneous current
flow a mechanical removing (by using the emery paper) of the
surface layer is important and was performed whenever it was
possible for our samples.

Particular care was taken to ensure the temperature stabilization.
The Hall voltage was measured at fixed temperatures for all three
pairs of Hall contacts to test and/or control the homogenous
current distribution through the sample and in field sweeps from
$-B_{\mathrm{max}}$ to $+B_{\mathrm{max}}$ in order to eliminate
the possible mixing of magnetoresistive components (due to the
misalignment of the opposite Hall contacts). Also, two contacts
along one side of the sample could be used as a pair of voltage
contacts to compensate (at zero field) the Hall signal,
  eliminating in that way the magnetoresistive component
directly from the measured data and thus  increasing the accuracy
of the Hall signal measurements. As it will be shown later, the
Hall resistivity is linear with applied fields up to 9 T. The Hall
coefficient $R_H$ is obtained as $R_H = (V_{xy}/IB)t$, where
$V_{xy}$ is Hall voltage determined as $[V_{xy}(B)-V_{xy}(-B)]/2$,
$I$ is the current through the crystal and $t$ is the sample
thickness. The presented $R_H$ values  are mean values (obtained
for 3 pairs of Hall contacts) and the error bars are the variance.

\section{Results}

\begin{figure}
\includegraphics*[scale=\myscale]{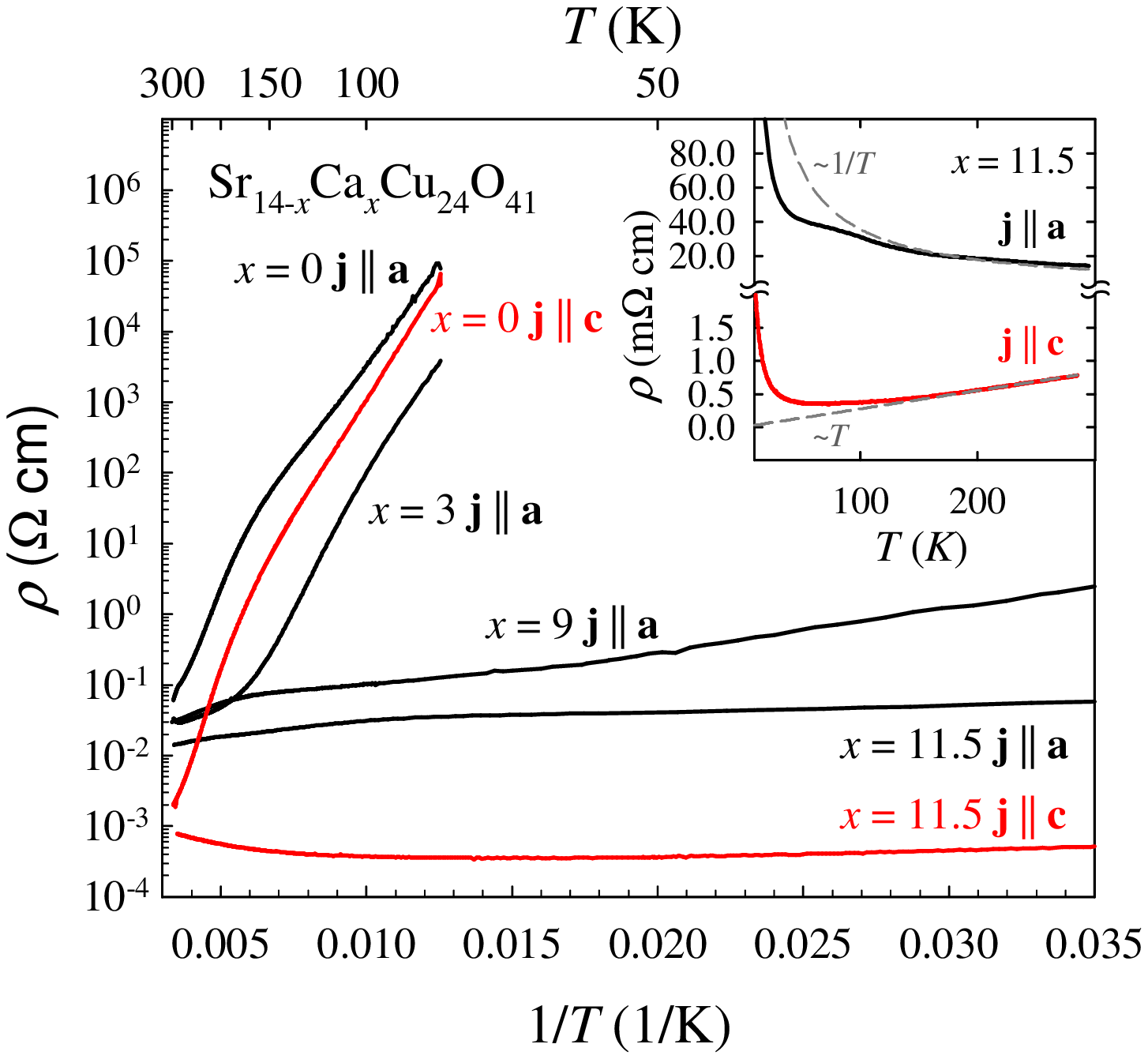}
\caption{\label{fig:1} (Color online) Temperature dependence of
the dc resistivity, $\rho$ vs.\ $1/T$ for different \sr\ compounds
measured along \textbf{a}-direction  \ja\  (black lines) and along
\textbf{c}-direction  \jc\  (red lines). Inset: Temperature
dependence $\rho$ vs.\ $T$ of the resistivity for $x=11.5$
compound for \ja\ and \jc. Dashed lines present related power
laws.}
\end{figure}

Figure \ref{fig:1} shows the dependence of the resistivity
$\rho(T)$ on inverse of temperature $1/T$, for \sr\ in the
temperature range $20\,\mathrm{K}<T<300\,\mathrm{K}$, measured on
samples with calcium part $x=0$, $x=3$, $x=9$ and $x=11.5$ along
\textbf{a} direction (\ja) and \textbf{c} direction (\jc). The
detailed resistivity measurements (from
$2\,\mathrm{K}-750\,\mathrm{K}$) on the same samples (except
$x=0$, \jc) were already done previously (for more details see
Refs.\ \onlinecite{Vule06,Vule03,vule05}). For our measurements
the samples were cleaned up, new contacts mounted and resistivity
re-measured. The same samples were used for Hall-effect
measurements as well. The room-temperature resistivity values for
\textbf{a} direction range from 70~m$\Omega$~cm for $x=0$ to
13~m$\Omega$~cm for $x=11.5$, and in \textbf{c} direction from
10~m$\Omega$~cm to 1~m$\Omega$~cm, which is in good agreement
with the previously published
data.\cite{Vule06,motoyama,Vule03,vule05} The presented results
refer to samples with homogeneous current flow and the same
samples were used for Hall-effect measurements as well.

The temperature dependence of resistivity, which is rather
different for various $x$ is in good agreement with the previously
published data.\cite{Vule06,motoyama,Vule03,vule05} For $x \leq
9$, $\rho_{a,c}(T)$ can be analyzed using a phenomenological law
for a semiconductor, $\rho \propto \exp[\Delta / T]$: for $x \leq
9$ all the samples are semiconducting already at room temperature
(for both \textbf{a} and \textbf{c} directions). The extracted high
temperature activation energies are $\Delta \sim 1000$~K for $x=0$
sample,  $\Delta \sim 500$~K   for $x=3$ and $\Delta \sim 120$~K
for $x=9$ (Refs.\ \onlinecite{Vule06,Vule03}). At lower
temperatures, the fact that for $x=0$ the activation energy below
about 150\,K appears to be lower for $\rho_{a}(T)$ than for
$\rho_{c}(T)$ is most probably sample dependent since we have used
two different samples for these measurements. The $x = 11.5$
sample shows a resistivity increase with decreasing temperature
for \textbf{a} direction and a metallic behavior in \textbf{c}
direction down to around 80\,K and semiconducting at lower
temperatures, as already known.\cite{motoyama,Vule03,vule05} In
the temperature region $140\,\mathrm{K}<T<300\,\mathrm{K}$,
$\rho_{a}(T)$ follows almost $T^{-1}$ while $\rho_{c}(T)$ follows
linear $T$ dependence (this different temperature variation is
seen the best in the inset of Fig.\ \ref{fig:1} where the dashed
lines present related power laws).

\begin{figure}
\includegraphics*[scale=\myscale]{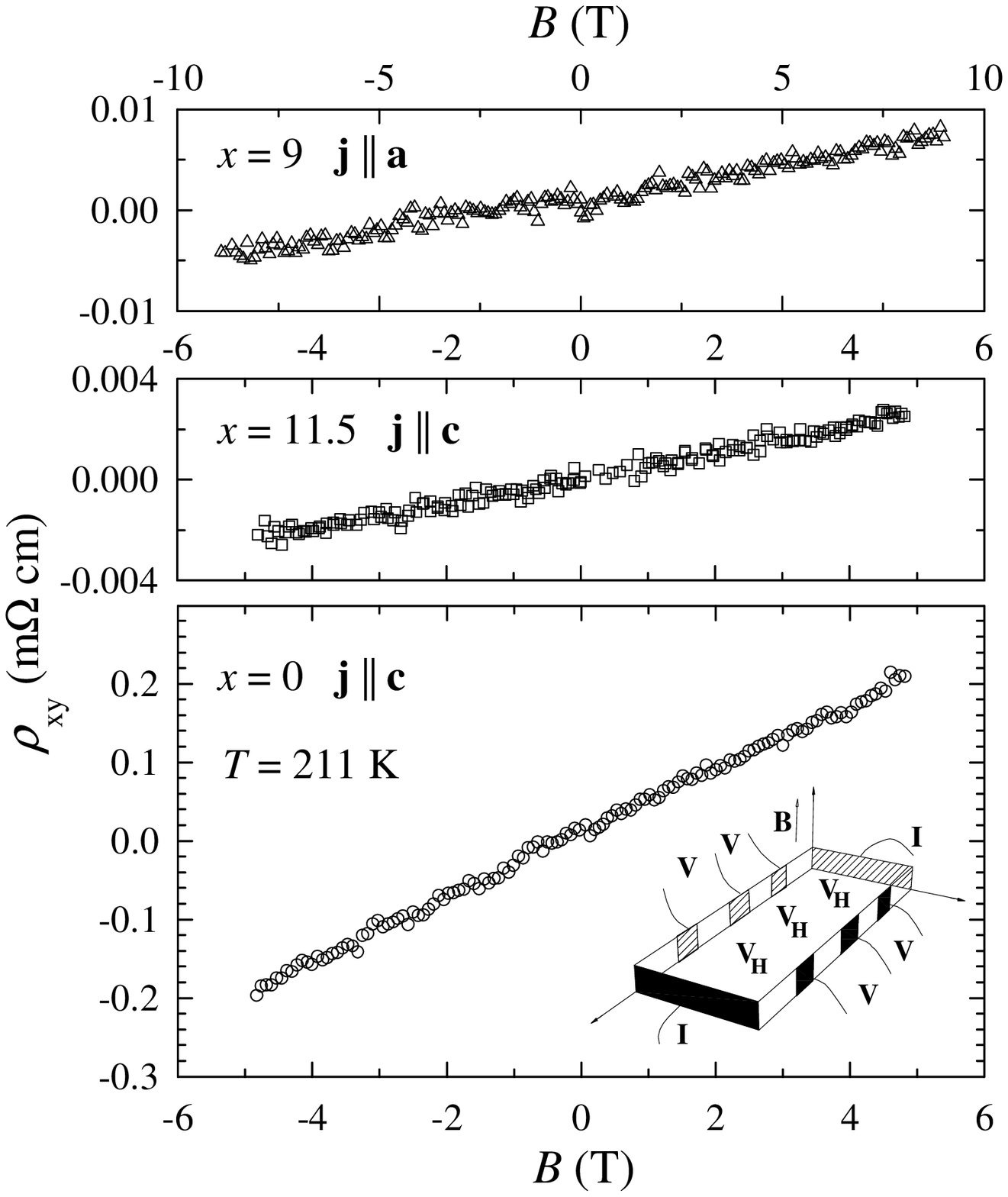}
\caption{\label{fig:2} Magnetic field dependence of the Hall
resistivity $\rho_{xy}$ of \sr\ for $x=0$ and 11.5 (\jc\ and \Bb~
up to 5 T) and for $x=9$ (\ja\ and \Bb~ up to 9 T); all at $T=211$
K. Also shown is the sample geometry and the arrangement of
contacts.}
\end{figure}

Figure \ref{fig:2} shows the magnetic-field dependence of the Hall
resistivity $\rho_{xy}$, at fixed temperature $T = 211$ K, for
three samples \sr\: $x=9$ (\ja), $x=11.5$ (\jc) and $x=0$ (\jc).
Similar $\rho_{xy}$ vs. $B$ sweeps at other fixed temperatures
show that the Hall resistivity is linear with field up to 9 T in
the whole temperature interval investigated and that the sign of
the Hall coefficient is positive.

For the \sr\ family, results from Hall effect
experiments for the x=12 compound at applied pressures of
0.3 and 1.0 GPa have been published.\cite{nakanishi} Our resistivity results
follow those published previously at ambient pressure. \cite{Vule06}

\begin{figure}
\includegraphics*[scale=\myscale]{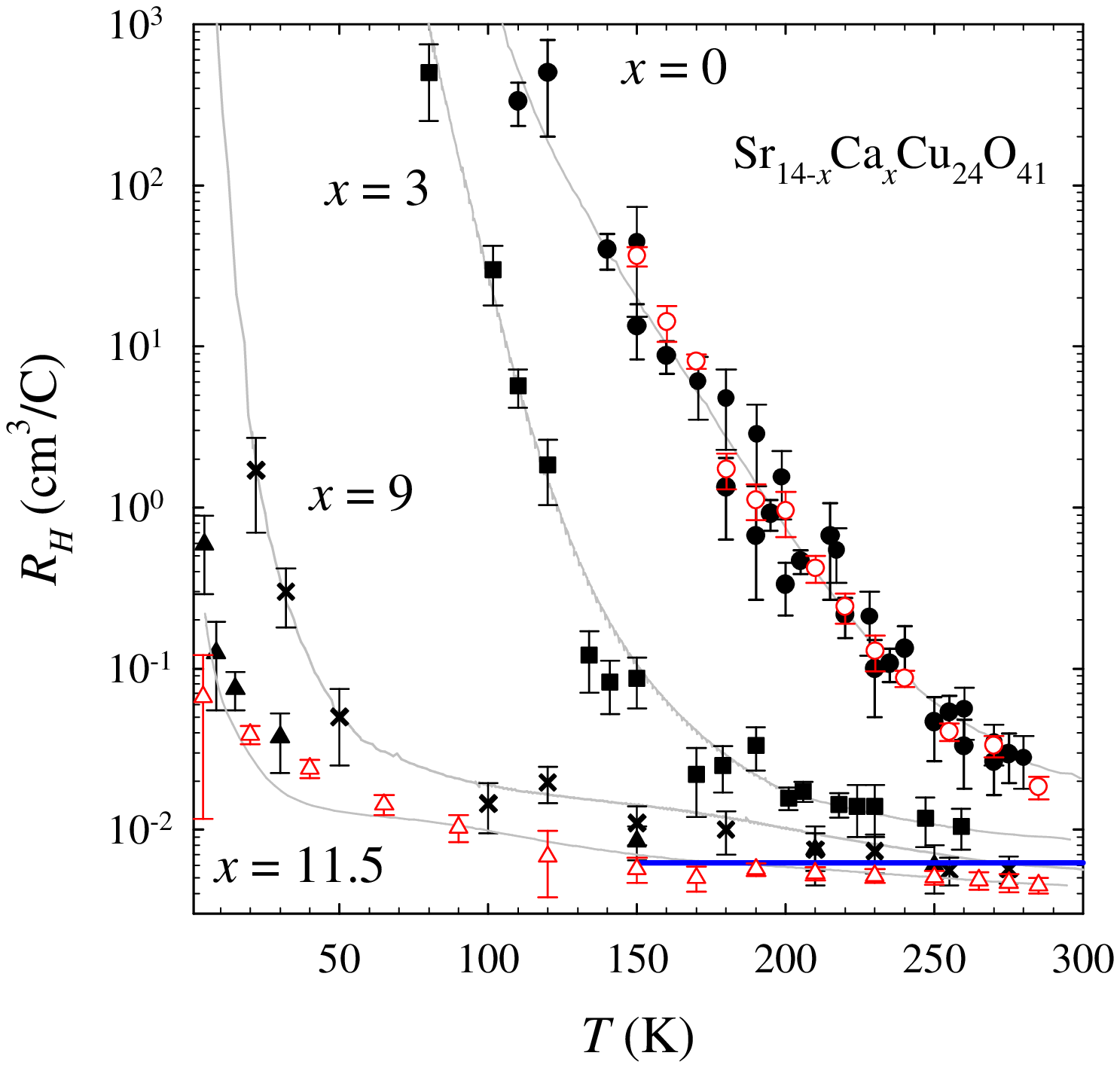}
\caption{\label{fig:3} (Color online) Temperature dependence of
the Hall coefficient $R_H$ for different \sr\ compounds. Full
symbols are for \ja\ geometry and empty symbols for \jc\ geometry.
The gray lines are scaled temperature dependences of resistivity
measured in \ja\ geometry. Solid  blue line is the calculated
value for the Hall coefficient (see text).}
\end{figure}

Figure \ref{fig:3} shows the temperature dependence of the Hall
coefficient $R_H$ for \sr\ in the temperature range
$4.2\,\mathrm{K}<T<300\,\mathrm{K}$, measured on samples with
calcium part $x=0$ (\ja, \Bb) (\jc, \Bb), $x=3$ (\ja, \Bb), $x=9$
(\ja, \Bb) and $x=11.5$ (\ja, \Bb) (\jc, \Bb). Each Hall
coefficient value was determined from one or more magnetic-field
sweeps at a fixed temperature.

The scaled temperature dependences of the resistivity for the
related samples (shown as gray lines in Fig. \ref{fig:3}) indicate
that the resistivity and the Hall coefficient follow the same
exponential law with the same activation energy. Furthermore, our
measurements do not show any anisotropy in the Hall effect: for
$x=0$ and  $x=11.5$ samples, the temperature dependences of $R_H$
are (within the experimental error) the same for (\ja, \Bb) and
(\jc, \Bb) geometry.  The measurements were done during cooling
and heating; the error bars indicate the changes in $R_H$ due to
very small temperature variations during magnetic field sweeps
(this particularly refers to the region where the resistivity
changes strongly with temperature).

The solid blue line in Fig.\ \ref{fig:3} represents the calculated
value of the Hall coefficient obtained assuming that five of the
six self-doped holes per f.u.\ are localized on the chains
and one of them is delocalized on the ladders.\cite{nucker}  In
this case   the carrier concentration is $\sim 10^{21}\,$cm$^{-3}$
which yields (by using the simple model) $R_{H}=1/ne= 6.25 \times
10^{-3}\mathrm{\ cm}^{3}/\mathrm{C}$ ($e$ is the electronic
charge). As shown, the room temperature Hall coefficient values
for all samples -- except the $x = 0$ one -- are close to this
calculated value. This issue will be discussed later. Another
feature, which will also be discussed later in more details, is
that for the $x=11.5$ compound the temperature variation of the
Hall coefficient (for both current directions) follows only the
\ja\ resistivity  (semiconducting) behavior.

\section{Discussion}

A  first comprehensive investigation of the anisotropic electrical
resistivity of \sr\   single crystals   at ambient pressure was
performed by Motoyama {\it et al.}\cite{motoyama}; later a number
of groups continued and extended these studies. The most detailed
dc electrical transport investigation of \sr\ single crystals
(with $x = 0$, 3, 6, 8, 9 and 11.5) was performed by Vuleti\'{c}
{\it et al.}\cite{Vule06} covering a large range of temperature
and all three crystallographic directions.  Our  resistivity
results follow these data closely. All the single crystals for
\jc\ and \ja\  (except \jc\ for $x = 11.5$) exhibit a rapid
increase in the resistivity upon cooling, following the activated
behavior characteristic for semiconductors. Calcium substitution
suppresses the activation energy (from 80 meV for $x=0$ to 9 meV
for $x=9$; see Ref.\ \onlinecite{Vule06} for more details). The value
of \va\ axis resistivity ($\rhoa$) is larger than the \vc\ axis
resistivity ($\rhoc$) by one to two orders of magnitude and for $0
\leq x < 11.5$  shows approximately the same activated behavior.
As a consequence, the anisotropy ratio ($\rhoa/\rhoc$) neither
depends strongly on $T$ at high temperatures nor varies
basically with Ca content.

As it was already shown,\cite{motoyama} the dc resistivity  for
$x=11.5$ exhibits a different behavior than all other $0 \leq x <
11.5$ compounds:  although at low temperatures, below $\approx
50$~K, the resistivity $\rhoc$ increases (as it does for  \ja\
direction, indicating a carrier localization), a metallic behavior
is seen  above 80 K. In other words, the $T$ dependence of $\rhoa$
(which is anomalous in the sense  that
$\mathrm{d}\rhoa/\mathrm{d}T<0$,
 i.e. semiconducting, in the temperature range where
$\mathrm{d}\rhoc/\mathrm{d}T>0$,  i.e. metallic), indicates a
non-coherent transport along the \va\ axis (along the rungs of the
ladders).  Moreover, an insulator-to-superconductor transition was
observed at $\sim 4.0$~GPa, accompanied by incoherent to coherent
crossover of the transverse (interladder, \va\ axis) charge
transport; i.e. $\rhoc$ and $\rhoa$ have shown quite similar
metallic temperature dependences, which indicates that both are
subject to the same scattering mechanism.\cite{nagata} The
application of pressure therefore triggers a dimensional crossover
in the charge dynamics from one to two, and the superconductivity
in this ladder compound might be a phenomenon in a 2D anisotropic
electronic system. Such findings indicate that, at ambient
pressure, charge dynamics is essentially one dimensional (1D) and
carriers are confined within each ladder.

The phenomenon of interlayer or interchain decoherence in
anisotropic metals is still poorly understood. In strictly 1D
systems all electronic states are known to be localized at $T=0$
in the presence of weak disorder, whereas in real materials with
finite interchain coupling $t_\perp$, the situation is different.
A considerable theoretical work has been devoted to this
subject.\cite{IshiguroBook98,hussey} Recently, in accord with
Prigodin and Firsov's\cite{prigodin} argumentation, that
 impurity scattering rates $\hbar/\tau_0 > t_\perp$ render the system
 effectively 1D and therefore susceptible to localization at low
 $T$, the influence of introduced disorder on anisotropic resistivity
 of PrBa$_2$Cu$_4$O$_8$ was analyzed and interpreted
 as a consequence of localization effects that are
observed for extremely small amounts of
disorder.\cite{narduzzoPRL} In other words, once the intrachain
scattering rate surpasses the interchain hopping rate, the
coherent interchain tunneling is suppressed, the system is
rendered effectively 1D and localization sets in. We believe that
this finding  shows qualitative agreement with our
 $\rho(T)$ data for $x=11.5$  (the metallic-like behavior in one direction
 only). In \sr\ an
incommensurability between the chains and ladders creates
distortions, which lead to additional modulations of the
crystallographic positions that are  intrinsic sources of
disorder. Moreover, there is an additional disorder introduced by
Ca-substitution.\cite{Vule06} Accordingly, the difference in  the
$\rhoc(T)$ and $\rhoa(T)$ behavior of $x=11.5$ compound may be
attributed to intrinsic disorder while the appearance of a
metallic behavior for $\rhoa(T)$ under pressure,\cite{nagata}
indicates that the interladder coupling starts to prevail over
one-dimensional effects, which dominate the charge transport at
ambient pressure.

Our results give a positive, hole-like, Hall coefficient which is
temperature dependent for all \sr\ samples. For two concentrations
($x=0$ and $x=11.5$) the measurements were performed in both
geometries \jc, \Bb\ and \ja, \Bb\ giving simi\-lar values (cf.\
Fig.\ \ref{fig:3}), thus indicating no anisotropy in $R_H(T)$.
According to Onsager's\cite{onsager} reciprocal relation for \Bb,
\RH\ measured in \ja\ and \jc\ geometry is the same unless the
time-reversal symmetry is broken. Since we observe that the
temperature dependence of \RH\ measured in both geometries is
approximately the same (the observed small  differences in
absolute Hall coefficient values are within experimental
uncertainties) we take that Onsager's relation is satisfied for
\sr\ and we comprehend both geometries as a good choice to
determine $R_H(T)$.

\begin{figure}
\includegraphics*[scale=\myscale]{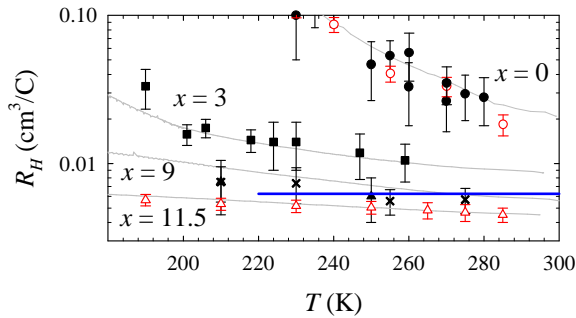}
\caption{\label{fig:4} (Color online) High temperature $R_H$ vs.
$T$, presented in more details in order to emphasize the values
around 300 K. }
\end{figure}

Turning now to the concentration variation of the Hall data, we
 first analyze the $x<11.5$ results. For $0 \leq x < 11.5$ both
$R_H(T)$ and $\rho(T)$ follow the same temperature dependences.
Their behavior is comparable to a conventional $p$-type
semiconductor, where at high enough temperature all doped holes
are activated and the effective number of carriers
$n_{\mathrm{eff}} \sim 1/e R_H$ saturates.\cite{smith} The Hall
coefficient increases with decreasing temperature; i.e.,
$R_H(T)\propto \exp [\Delta/T]$ is thermally activated and the
activation energy $\Delta$ for $x<11.5$ agrees well with that
obtained from the resistivity data.

For $x = 0$ (and possibly $x=3$) $n_{\mathrm{eff}}$ is not fully
saturated at 300~K (because $\Delta \gg T$), and the actual hole
number is  higher.

\begin{figure}
\includegraphics*[scale=\myscale]{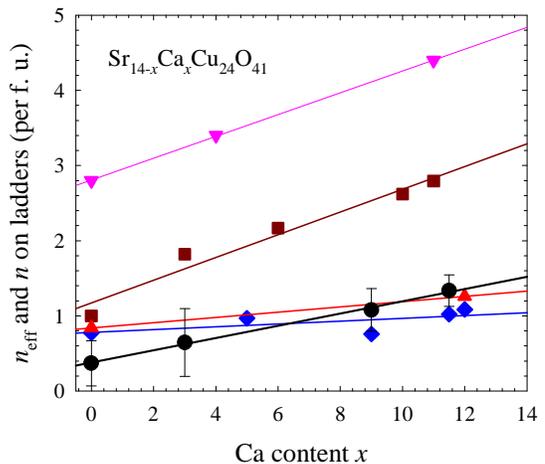}
\caption{\label{fig:5}  (Color online) The effective number of
carriers $n_{\mathrm{eff}}=V/(4eR_{H})$ ($\bullet$  our data) and
the number of holes $n$ per formula unit in the ladder vs. calcium
content $x$ in \sr.(Data taken from $\blacksquare$  Osafune et
al.~\cite{osafune},$\blacklozenge$ N\"{u}cker et
al.~\cite{nucker}, $\blacktriangle$  Piskunov et
al.~\cite{piskunov}, $\blacktriangledown$ Rusydi et
al.~\cite{rusydi})}
\end{figure}

From the measured  Hall coefficient \RH\ values for different
Ca-substitutions (Fig.\ \ref{fig:4}) we can calculate the
effective  number of carriers $n_{\mathrm{eff}}=V/(4eR_{H})$ per
f.u.\ at room temperature ($V$ is the volume of the unit
cell, $e$ is the electronic charge and the factor 4 in the
denominator describes that the unit cell contains four formula
units; the changes in the unit cell volume with Ca content have
been taken into account\cite{siegrist,mccarron,isobe}). For
$x=11.5$ we consider that $R_H(T)$ values at temperatures that
approach room temperature do not change significantly (and we
calculate $n_{\mathrm{eff}}$ in an equivalent way as for
$x<11.5$); the related temperature dependence will be discussed
more in the last part of this section. The $n_{\mathrm{eff}}$
values are displayed in Figure \ref{fig:5}. The same figure  also
shows the number of holes per f.u.\ ($n$) in \sr\ ladders
for $0 \leq x \leq 12$ obtained by different experimental
techniques.\cite{osafune,nucker,piskunov,rusydi}

Comparing our $n_{\mathrm{eff}}$ values with $n$ obtained from
other experiments\cite{osafune,nucker,piskunov,rusydi} we find a
 good agreement with NEXAFS and NMR results,\cite{nucker,piskunov}
while optical \cite{osafune} and resonant soft x-ray scattering
\cite{rusydi} data give  much higher values, and at this point  we
cannot clarify the origin of this discrepancy.  Our measurements
yield  for the  $n_{\mathrm{eff}}(x=11.5)- n_{\mathrm{eff}}(x=0)$
a value  which agrees with Refs.\ \onlinecite{nucker,piskunov}
bearing in mind the inaccuracy in the determination of
$n_{\mathrm{eff}}(x=0)$. We believe that our result describes well
the fact that a minor change in number of carriers on the ladders
is responsible for a pronounced change in resistivity with Ca
doping. Here we compare our results of $R_H$ for $x=11.5$,
especially around room temperature, where $R_H \approx 4.5
\times10^{-3}\,$cm$^{-3}/$C (that gives $n_{\mathrm{eff}} \approx
1.3$ per f.u.\ in the ladder) with the published data on
temperature dependence of Hall coefficient for $x=12$ at pressures
of 0.3 and 1.0 GPa (Ref.\ \onlinecite{nakanishi}), which give
around room temperature $R_H \approx 2
\times10^{-3}\,$cm$^{-3}/$C. The considerable reduction in $R_H$
around room temperature under pressure confirms the NMR
data\cite{piskunov} where it was suggested that the important role of
high pressure for reaching conditions for the stabilization of
superconductivity in Sr$_{2}$Ca$_{12}$Cu$_{24}$O$_{41}$ is an
increase of the hole density in the ladder layers. Note finally
that it was also suggested \cite{piskunov} that the hole
distribution is temperature dependent and that the back-transfer
of holes from the ladders into the chains takes place gradually
with decreasing temperature, but we cannot speculate about that
from Hall effect data.  To summarize this part of the discussion,
we can point out that the effective number of carriers
$n_{\mathrm{eff}}$ corresponds to the number of holes per f.u.\ in the
 ladder obtained by other important experimental techniques. Assuming
that Cu-O chains do not contribute to the Hall effect (the holes
on the chains are localized\cite{Vule06}) we can conclude that
$n_{\mathrm{eff}}$ denote the holes in the ladders that
participate in dc transport.

\begin{figure}
\includegraphics*[scale=\myscale]{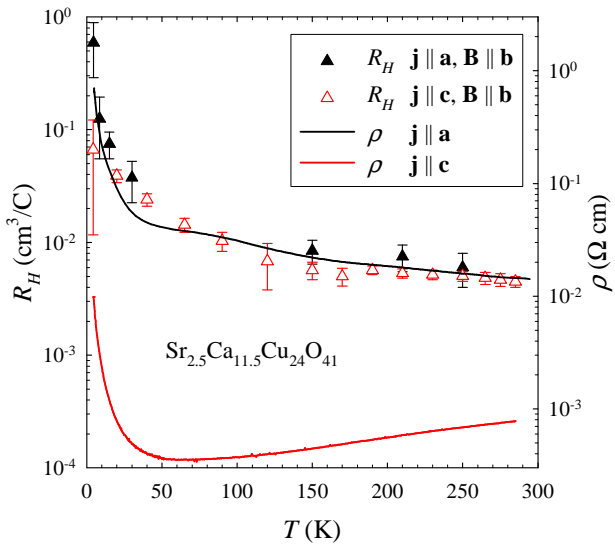}
\caption{\label{fig:6} (Color online) Temperature dependence of
the Hall coefficient $R_H$ and resistivity $\rhoc$ and $\rhoa$ for
the $x=11.5$ compound of \sr. Full symbols are $R_H$ for \ja\
geometry and empty symbols for \jc\ geometry. Red line:   $\rhoc$,
black line: $\rhoa$.}
\end{figure}

Our results for $x=11.5$  are quite different.
Figu\-re~\ref{fig:6} shows in more detail the temperature
dependence of $\rhoc(T)$, $\rhoa(T)$ and the Hall coefficient
$R_H(T)$ in both \ja, \Bb\ and \jc\, \Bb\ geometry (as already
pointed out and presented in Fig.~\ref{fig:6}, the Hall
coefficient is isotropic). Here we should mention that the
published data on pressure dependence of Hall coefficient for
$x=12$ at pressures of 0.3 and 1.0 GPa, have shown different
temperature dependences for $R_H(T)$ (\jc\, \Bb) and
$\rhoc(T)$.\cite{nakanishi} Our results show that although the
$R_H(T)$ values do not differ for two different current
directions, the temperature variation in $R_H(T)$ follows that of
$\rhoa(T)$. The different temperature variation in  $R_H(T)$ and
$\rhoc(T)$ is particularly evident above 140\,K: while $\rhoc(T)$
increases linearly with temperature, the Hall coefficient is
proportional to $1/T$.

Such a behavior is typical for high temperature copper-oxide
superconductors, albeit in a broad temperature interval, which
metallic state is characterized by unusual and distinct
temperature dependences in the transport
properties\cite{hussey,DagottoHTC} that deviate from the
conventional Fermi-liquid behaviors. Optimally-doped cuprates are
characterized by a linear-$T$ resistivity that survives for all $T
> T_c$, while the in-plane Hall coefficient varies approximately
as $1/T$ over a wide temperature range.\cite{carrington, hwang}
The $R_H$ of the hole-doped high-$T_c$ cuprates varies markedly
with both the number of holes doped onto the copper oxide planes
and temperature. The inverse Hall angle $\cot \theta_H =
\rho_{ab}/R_H B$ shows a quadratic $T$-dependence over a
remarkably broad temperature range and holds for a wide range of
doping in most cuprates\cite{ono,hussey} ($\rho_{ab}$ denotes
in-plane resistivity). This unconventional behavior has led
theorists to develop a number of models with different approaches,
such as two-lifetime picture of Anderson,\cite{anderson} marginal
Fermi-liquid phenomenology\cite{varma} and models based on
fermionic quasiparticles that invoke specific (anisotropic)
scattering mechanisms within the basal
plane.\cite{husseyEPJB,carrington,
castellani,ioffe,zheleznyakHTC,jawad}  Currently, the overall
experimental situation does appear to support models in which
anisotropic linear-$T$ scattering, in conjunction with the Fermi
surface curvature and band anisotropy, is primarily responsible
for the $T$-dependence of $\rho_{ab}$ and $R_H(T)$. However, the
origin of the anisotropic scattering is not known at
present,\cite{hussey} and new interpretations of the normal state
transport properties of high-$T_c$ are expected.

\begin{figure}
\includegraphics*[scale=\myscale]{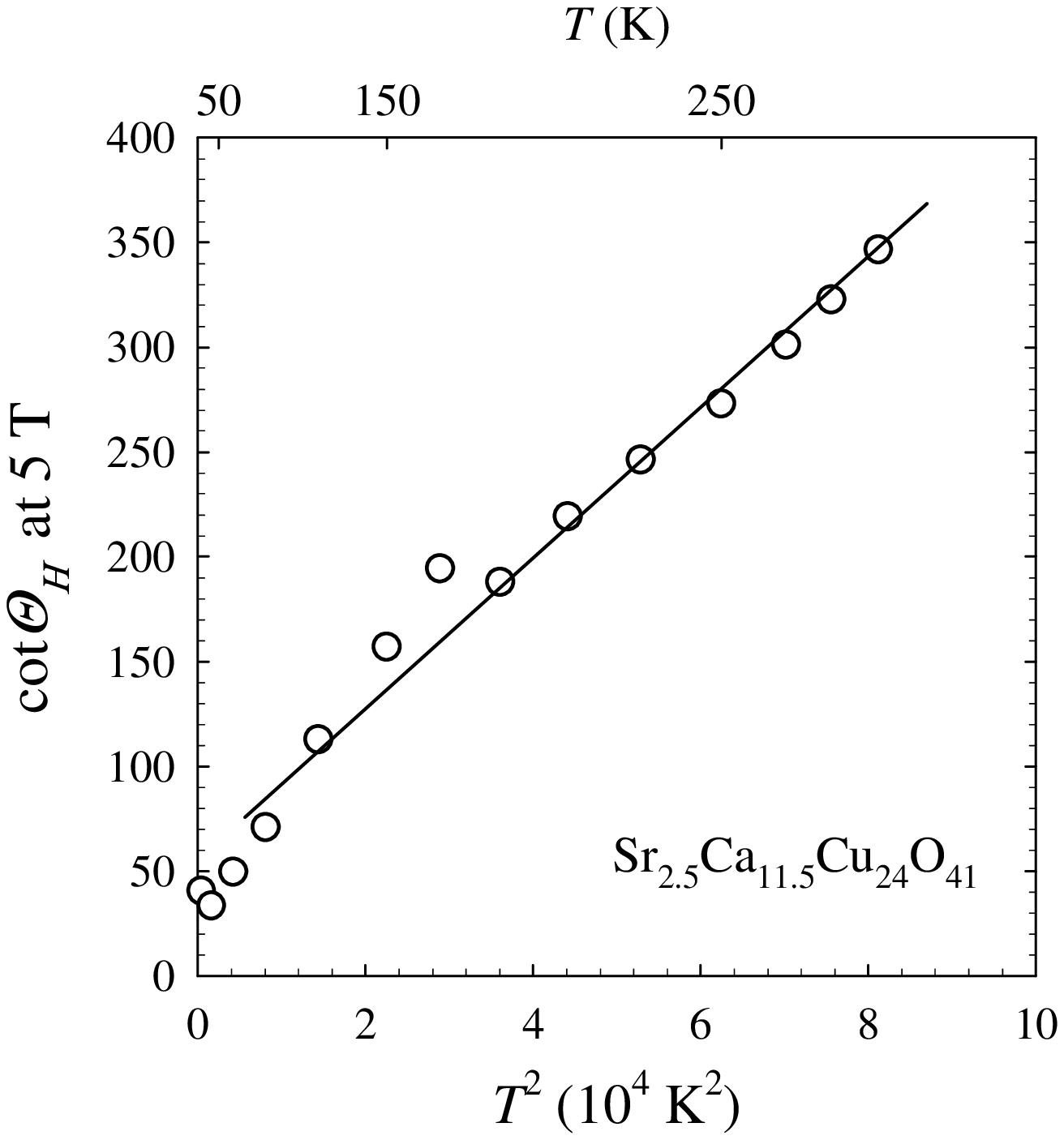}
\caption{\label{fig:7} $T^2$  dependence of  $\cot \theta_H$ (at 5
T) for $x=11.5$ in \sr.}
\end{figure}

Figure~\ref{fig:7} shows $\cot \theta_H = \rho_{c}/R_H B$ (at 5 T)
vs.\ $T^2$ for $x=11.5$ compound, and it is   clearly perceived
that for $T>140\,$K the inverse Hall angle follows $T^2$
dependence up to room temperature. Since the unusual temperature
dependences of $\rho_{ab}$ and $R_H(T)$ of high temperature
copper-oxide superconductors are frequently ascribed to the
intrinsic property of the CuO$_2$ planes, it is quite interesting
that similar behavior is now found for highly anisotropic
Cu$_2$O$_3$ ladder plane. The \abplane\ plane in high-$T_c$ is
almost isotropic, with metallic-like resistivities for both
perpendicular directions. On the other hand,
 for \sr\ and $x=11.5$ the anisotropy ratio $\rhoa /\rhoc$  is around
13 at room temperature and becomes enhanced at lower temperatures
reaching about 55 at 140\,K  (due to metallic-like $\rhoc(T)$ and
non-metallic like $\rhoa(T)$ variations). This leads to the
interesting conclusion:   the well known $T^2$ dependence of the
inverse Hall angle, which seems to be largely doping independent
in cuprates\cite{koki,ando04} is not changed by the increased
anisotropy in the ladder plane. Further investigations comprising
pressure dependence of $\rhoc(T)$, $\rhoa(T)$ and $R_H(T)$ for
different pressures (up to the pressures when superconductivity
occurs) and different $x$ may give more information about the
normal state properties of \sr\ and could be an important step
towards a correct microscopic theory of cuprate superconductivity.

\section{Conclusion}

In summary, we have reported measurements of the Hall coefficient
$R_H(T)$ at ambient pressure of the quasi-one dimensional cuprate
\sr\ for $0 \leq x \leq 11.5$. It is known that isovalent
Ca-substitution does not change the total hole count, but causes
additional transfer of holes from the chains (where they are
localized) into the ladders (where they are rather mobile), but as
far as the amount of transferred holes is concerned, the
experiments which have been performed up to now, have given
contradictory results.

Our findings give a positive, holelike, Hall coefficient which is
temperature dependent for all samples. For $x<11.5$ the Hall
coefficient is activated, and  the activation energy   agrees well
with that obtained from the resistivity data. The observed
behavior is that of a conventional $p$-type semiconductor. For
$x=11.5$ $\rhoc(T)$ and $R_H(T)$ follow different temperature
dependences: above 140\,K $\rhoc(T)$ increases linearly with
temperature, while the Hall coefficient is inversely proportional
to $T$, giving an overall $T^2$ dependence of the inverse Hall
angle $\cot \theta_H$ for $140\,\mathrm{K} < T < 300\,\mathrm{K}$.
Such behavior, which is well known and typical  for high
temperature copper-oxide superconductors,  seems independent of
the pronounced anisotropy in the Cu$_2$O$_3$ ladder plane.

A comparison of our estimate for the effective number of carriers
$n_{\mathrm{eff}} \sim 1/R_H$ for \sr\ ($0 \leq x \leq 11.5$) at
300\,K with the numbers of holes in ladders (obtained by different
experimental techniques) shows good agreement with NEXAFS and
NMR results and indicates that $n_{\mathrm{eff}}$ corresponds to
the holes in the ladders that participate in dc transport. The
difference $n_{\mathrm{eff}}(x=11.5)-n_{\mathrm{eff}}(x=0) = 0.96$
(this value could be  even lower due to, probably underestimated,
$n_{\mathrm{eff}}$  for $x=0$) leads us to conclude that a minor
change in number of carriers on the ladders is responsible for a
pronounced change in resistivity with Ca substitution.

\begin{acknowledgments}
We acknowledge the support of the Croatian Ministry of Science,
Technology and Sports under project No.\ 035-0000000-2836 and No.\
119-1191458-1023.
\end{acknowledgments}

\end{document}